\documentclass[aps,letterpaper,twocolumn,pra,footinbib,floatfix,showpacs]{revtex4}
\usepackage{amsmath,graphicx,natbib,feynmp,psfrag}

\begin{document}
\author{Bhuvanesh Sundar}
\affiliation{Laboratory of Atomic and Solid State Physics, Cornell University, Ithaca New York 14850, USA}
\author{Erich J. Mueller}
\affiliation{Laboratory of Atomic and Solid State Physics, Cornell University, Ithaca New York 14850, USA}
\title{Proposal to directly observe the Kondo effect through enhanced photoinduced scattering of cold fermionic and bosonic atoms}
\date{\today}

\begin{abstract}
We propose an experimental protocol to directly observe the Kondo effect by scattering ultracold atoms. We propose using an optical Feshbach resonance to engineer Kondo-type spin-dependent interactions in a system with ultracold $^6$Li and $^{87}$Rb gases. We calculate the momentum transferred from the $^{87}$Rb gas to the $^6$Li gas in a scattering experiment and show that it has a logarithmically enhanced temperature dependence, characteristic of the Kondo effect and analogous to the resistivity of alloys with magnetic impurities. Experimentally detecting this enhancement will give a different perspective on the Kondo effect, and allow us to explore a rich variety of problems such as the Kondo lattice problem and heavy-fermion systems.
\end{abstract}

\newcommand{\ha}{\hat{a}}
\newcommand{\hb}{\hat{b}}
\newcommand{\+}{^\dagger}
\newcommand{\ua}{\uparrow}
\newcommand{\down}{\downarrow}
\newcommand{\hH}{\hat{H}}
\newcommand{\HInt}{\hat{H}_{\rm{int}}}
\newcommand{\Li}{$^6$Li}
\newcommand{\Rb}{$^{87}$Rb}

\maketitle

\section{Introduction}
Ultracold atomic gases provide a platform to engineer model Hamiltonians relevant for condensed matter physics phenomena. One such intriguing phenomenon is the Kondo effect \cite{Kondo,hewson}. In this paper we propose an experimental protocol to engineer and measure the scattering properties of Kondo-like interactions between ultracold atoms. Such an experiment would give a new perspective on an iconic problem.

The Kondo effect is a transport anomaly that arises when itinerant electrons have spin-dependent interactions with magnetic impurities. The source of the phenomenon is a spin-singlet many-body bound state formed between the Fermi sea and an impurity. This bound state leads to resonant scattering of itinerant electrons off the screened impurities. As the temperature is lowered, this resonant scattering dominates over other scattering processes and leads to a characteristic logarithmic temperature dependence of the resistivity of the material. When the interactions between the electrons and the impurity are spin independent, no such bound state is formed, and the scattering is not enhanced.

Despite intense research, some questions about the Kondo effect remain unresolved and some of the key theoretical predictions have never been directly seen. For example, the electron cloud which screens the spin on the impurity has never directly been imaged \cite{affleckArXivKondo,affleckKondoScreeningReview,KondoCloud1,KondoCloud2}. More importantly the analogous problem with an array of interacting impurities (the Kondo lattice) has aspects which are not well understood \cite{KLMReview}. Exploring the Kondo lattice problem is of paramount importance to the understanding of heavy fermion systems and quantum criticality \cite{heavyFermionsReview,heavyFermionsReview2}.

In this paper we propose using cold atoms to directly observe enhanced Kondo scattering. We envision a system consisting of a spin-1/2 Fermi gas and a dilute Bose gas with spin $S$, where bosonic atoms play the role of magnetic impurities and fermionic atoms play the role of electrons. To strengthen the analogy with immobile spin impurities in the Kondo model, we consider bosons which are much heavier than the fermions. Fermion-boson pairs such as \Li-\Rb\, $^7$Li-$^{85}$Rb or \Li-$^{133}$Cs are good candidates with large mass ratios. Alkaline-earth-metal and rare-earth atoms are also promising.

\begin{figure}[t]
 \hspace{-3cm}\includegraphics[width=1.3\columnwidth]{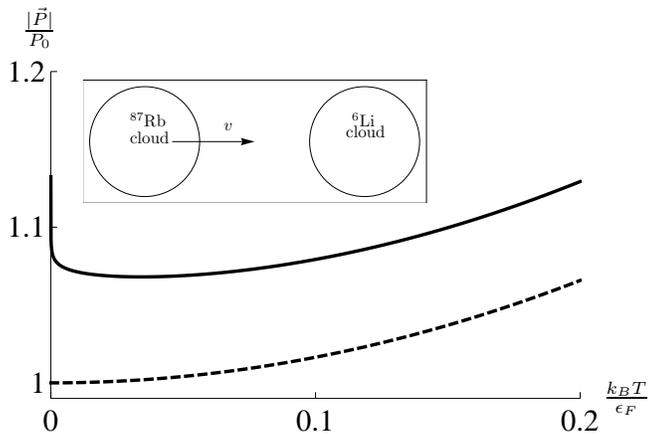}
 \caption{Temperature dependence of the momentum $\vec{P}$ transferred from bosons to fermions in a scattering experiment with photoinduced interactions. Both $\vec{P}$ and temperature have been rescaled to dimensionless quantities. $P_0$ denotes the momentum transferred to the Fermi gas at zero temperature when the interactions are spin independent ($g_s=0$). (Solid line) Spin-dependent interactions between spin-1/2 fermions and spin-1 bosons with $g_s=\frac{-1}{3}g_n = \frac{0.1\epsilon_F}{N/v}$; (Dashed line) Spin-independent interactions ($g_s=0$ and $g_n=-3\times\frac{0.1\epsilon_F}{N/v}$). The minimum in $\vec{P}$ is a signature of the Kondo effect, and may be detected experimentally. In Sec. \ref{sec:engineering interaction} we estimate experimental parameters to achieve the interaction strength used here. Inset shows a cartoon of the collision.
 }
 \label{fig:temperature plot}
\end{figure}

We consider a rotationally symmetric interaction between the ultracold atoms, which includes both density-density and spin-dependent interactions. We present an experimental protocol to produce such an interaction using an optical Feshbach resonance. For this general interaction, we calculate that the scattering cross section is strongly enhanced by the Kondo effect. We propose directly measuring this enhancement by launching the Bose gas into the Fermi gas with a small velocity. One would then measure the momentum transferred to the Fermi gas. A number of related experiments have been used to probe atomic scattering in the past \cite{SpielmanScattering,thomas2004imaging,gensemer2012direct,robins2008probing}. We show that at temperatures smaller than the Fermi temperature, the final momentum of the Fermi gas varies logarithmically with temperature, analogous to the resistance of electrons in an alloy with magnetic impurities. The temperature dependence of the transferred momentum, depicted in Fig. \ref{fig:temperature plot}, has a minimum which is a signature of the Kondo effect, and this minimum can be detected at experimentally accessible temperatures. Alternatively, the enhanced scattering could be seen in the damping of collective modes of the atomic clouds in a trap \cite{collectiveOscillations}.

This paper is organized as follows. In Sec. \ref{sec:model} we introduce our atomic system and the model we consider. In Sec. \ref{sec:engineering interaction} we explain how an optical Feshbach resonance can be used to produce the interactions considered in our model. In Sec. \ref{sec:scattering} we calculate the momentum exchanged in a scattering experiment between atomic clouds. We calculate the momentum transferred as a function of temperature perturbatively up to third order in the interaction strength. We explicitly describe all parts of our calculation in the appendix. We summarize in Sec. \ref{sec:summary}.

\section{Model}\label{sec:model}
In this section we describe our model. In Sec. \ref{sec:engineering interaction} we describe how to experimentally implement our model.

We build our system out of spin-1/2 fermions and spin-$S$ bosons. In our implementation these will be hyperfine spins. We let the operators $\hat{\tilde a}_{r\alpha}\+$ and $\hat{\tilde b}_{r\mu}\+$ create fermionic and bosonic atoms at position $\vec{r}$ and spin projection $\alpha=\ua,\down$ or $\mu=-S,..,S$ along the $z$-axis. Their Fourier transforms,
\begin{equation}\begin{split}
 \ha_{k\alpha}\+ &= \frac{1}{\sqrt{V}}\int d^3\vec{r}\ \hat{\tilde a}_{r\alpha}\+ e^{i\vec{k}\cdot\vec{r}},\\
 \hb_{k\mu}\+ &= \frac{1}{\sqrt{V}}\int d^3\vec{r}\ \hat{\tilde b}_{r\mu}\+ e^{i\vec{k}\cdot\vec{r}},
\end{split}\end{equation}
create particles in momentum eigenstates. Above, $V$ is the volume of the system.

We explore a model with a Hamiltonian $\hH = \hH_0 + \HInt$. The first term models the kinetic energy of the fermions and bosons,
\begin{align}
 &\hH_0 = \frac{V}{(2\pi)^3}\int d^3\vec{k} \left(\sum_\alpha(\epsilon_k-\mu)\ha_{k\alpha}\+\ha_{k\alpha} + \sum_\mu E_k\hb_{k\mu}\+\hb_{k\mu}\right),\notag\\
 &\epsilon_k = \frac{\hbar^2k^2}{2m_a},\ E_k = \frac{\hbar^2k^2}{2M_b}.
\end{align}
For the interactions modeled by $\HInt$, we consider a generic form of local spherically symmetric pairwise Bose-Fermi interactions. Since the fermions have spin-1/2, the most general such interaction has the form,
\begin{equation}
 \HInt = \int d^3\vec{r}\sum_{\alpha\beta\mu\nu} \hat{\tilde a}_{r\alpha}\+\hat{\tilde a}_{r\beta}\hat{\tilde b}_{r\mu}\+\hat{\tilde b}_{r\nu} \left( g_s\vec{\sigma}^{(a)}_{\alpha\beta}\cdot\vec{\sigma}^{(b)}_{\mu\nu} + g_n\delta_{\alpha\beta}\delta_{\mu\nu} \right).
 \label{eqn:HKondoRealSpace}
\end{equation}
We denote the vector of spin matrices for the fermions and bosons by $\vec{\sigma}^{(a)}$ and $\vec{\sigma}^{(b)}$, and $\delta$ refers to the Kronecker delta function. It is important to note that $\hH_{\rm{int}}$ contains terms where $\alpha\neq\beta$ and $\mu\neq\nu$. This encodes the fact that the atoms exchange spin when they collide. We point out that spherical symmetry of the Hamiltonian is not a necessary feature to observe Kondo physics. Any Hamiltonian which allows spin exchange processes at third order of interaction strength would produce an enhanced scattering cross section at low temperatures. We restrict ourselves to interactions modeled by Eq. (\ref{eqn:HKondoRealSpace}), and we show in Sec. \ref{sec:engineering interaction} that this has a simple experimental realization.

It is useful to rewrite $\HInt$ in momentum space as
\begin{align}
 &\HInt = \frac{V^2}{(2\pi)^9}\int d^3\vec{k}\int d^3\vec{p}\int d^3\vec{q} \sum_{\alpha\beta\mu\nu}\label{eqn:HKondo}\\
 & \ha_{k+q,\alpha}\+\ha_{k+p,\beta} \hb_{k-q,\mu}\+\hb_{k-p,\nu} \left( g_s\vec{\sigma}^{(a)}_{\alpha\beta}\cdot\vec{\sigma}^{(b)}_{\mu\nu} + g_n\delta_{\alpha\beta}\delta_{\mu\nu} \right)\notag.
\end{align}
Our model in Eq. (\ref{eqn:HKondo}) differs from the one in the spin-$S$ Kondo model \cite{Kondo} in two respects. The bosonic atoms, which play the role of impurities, are mobile. Due to their large mass however, the recoil of the bosonic atoms can be neglected, and formally the physics is equivalent to that of immobile spin impurities. In addition to the regular spin-$S$ Kondo-like interaction, Eq. (\ref{eqn:HKondo}) contains a density-density interaction. We show that in spite of such an additional interaction term, the momentum transferred to the Fermi gas in a scattering experiment still has a minimum at a certain temperature, albeit at a lower temperature than the case with no density-density interaction.

The interaction we have considered in Eq. (\ref{eqn:HKondoRealSpace}) does not occur in typical cold atom experiments in which interaction strengths are tuned using a magnetic Feshbach resonance. In a typical magnetic Feshbach resonance, spin-exchange collisions are off-resonance and will not be observed. In the following section we propose using an optical Feshbach resonance to produce the interaction in Eq. (\ref{eqn:HKondoRealSpace}) .

\section{An experimental setup}\label{sec:engineering interaction}
In this section we describe our proposal to experimentally implement the model introduced in Sec. \ref{sec:model} using \Li\ and \Rb\ atoms as our itinerant fermions and spin impurities. As we will show, producing a strong interaction between \Li\ and \Rb\ using an optical Feshbach resonance requires a large matrix element for photoassociation. Experiments \cite{opticalFeshbachResonances1} show $^7$Li and $^{85}$Rb to have the highest photoassociation rate coefficient among all the bialkali metal combinations. We expect their isotopes \Li\ and \Rb\ to have similar photoassociation rates, and we chose \Li\ and \Rb\ in our proposal to produce the Kondo model because they are readily available in ultracold atomic experiments. The \Li\ and \Rb\ atoms have quantum numbers $S=1/2$, $L=0$ and $I=1$ and $3/2$.

In an optical Feshbach resonance, a laser beam provides a coupling between the open scattering channel and a closed channel containing a bound state \cite{fedichevOFR,OFRreview,ChengChinReview,theisOFRlosses}; here the open channel is an electronic spin-singlet of \Li\ and \Rb, and the bound state is a highly excited LiRb molecular state. When the laser is far detuned from resonance with the bound state, the bound state can be adiabatically eliminated, and we are left with an AC Stark shift for the \Li-\Rb\ singlet. The triplet state sees no Stark shift. This provides a mechanism for spin exchange. While this optically induced spin exchange has not yet been experimentally observed, there have been extensive studies of both elastic and inelastic scattering properties near heteronuclear optical Feshbach resonances of $^7$Li and $^{85}$Rb \cite{opticalFeshbachResonances1,opticalFeshbachResonances2}. Thus the transition frequencies for forming $^7$Li$^{85}$Rb molecules are well known. We expect that the linewidths, transition matrix elements, and spectral densities for other alkali-metal combinations such as \Li\Rb\ molecules will be similar.

Below we provide a mathematical framework to model the optical Feshbach resonance and obtain an effective interaction between the \Li\ and \Rb\ atoms. All the physics described in this section is local, and we have dropped the index labeling the position of the atoms from the second-quantized operators.

The energy density for the relevant electronic and nuclear degrees of freedom in each atom and molecule is of the form,
\begin{equation}
 \hat{\tilde H} = \hH_{\rm{HF}}^{\rm{Li}} + \hH_{\rm{HF}}^{\rm{Rb}}  + \hH_{\rm{mol}} + \hH_{\rm{Fesh}}.
\end{equation}

$\hH_{\rm{mol}}$ models the binding energy of the molecule:
\begin{equation}
 \hH_{\rm{mol}} = \sum\limits_{mm'} E_b\hat{\gamma}_{mm'}^\dagger\hat{\gamma}_{mm'},
\end{equation}
where $\hat{\gamma}^\dagger_{mm'}$ creates a molecule with an electronic spin $S=0$ and electronic orbital angular momentum $J=1$. The indices $m$ and $m'$ label the nuclear spins of the \Li\ and \Rb\ atoms. If the quantization axis of the electronic orbital angular momentum is chosen along the direction of angular momentum of the laser photon inducing the Feshbach resonance, then only one of the molecular states in the $J=1$ triplet is coupled via the laser to the atomic singlet. We denote the binding energy of this molecular state by $E_b$.

The hyperfine Hamiltonians for the atoms are
\begin{equation}\begin{split}
 \hH_{\rm{HF}}^{\rm{Li}} &= hA_{\rm{Li}} \sum\limits_{\substack{m_S,m_S'\\m_I,m_I'}} \ha_{m_Sm_I}\+\ha_{m_S'm_I'} \vec{\sigma}_{m_Sm_S'}^{(1/2)}\cdot\vec{\sigma}_{m_Im_I'}^{(1)}\\
 \hH_{\rm{HF}}^{\rm{Rb}} &= hA_{\rm{Rb}} \sum\limits_{\substack{m_S,m_S'\\m_I,m_I'}} \hb_{m_Sm_I}\+\hb_{m_S'm_I'} \vec{\sigma}_{m_Sm_S'}^{(1/2)}\cdot\vec{\sigma}_{m_Im_I'}^{(3/2)}
\end{split}\end{equation}
where $h$ is Planck's constant, $A_{\rm{Li}} = 152\rm{MHz}$ and $A_{\rm{Rb}} = 3.4\rm{GHz}$ are the hyperfine coupling constants of \Li\ and \Rb \cite{hyperfineCouplingConstants}, $\vec{\sigma}^{(S)}$ is the vector of spin-$S$ matrices, and $\ha_{m_Sm_I}\+$ and $\hb_{m_Sm_I}\+$ create a \Li\ and \Rb\ atom in the state $|m_Sm_I\rangle$. In terms of the hyperfine eigenstates,
\begin{equation}
 |m_S,m_I\rangle = \sum_{F,m_F}C_{m_Sm_I}^{Fm_F} |F,m_F\rangle \label{eqn:CGcoeffs}
\end{equation}
where $C_{m_Sm_I}^{Fm_F}$ are Clebsch-Gordan coefficients.

The terms in $\hH_{\rm{Fesh}}$ describe the interactions between the photoassociation laser and the atoms. We model this photoinduced molecular formation by
\begin{align}
 \hH_{\rm{Fesh}} = &\sum_{mm'} \Omega e^{i(\vec{k}\cdot\vec{r}-\omega t)} \hat{\gamma}_{mm'}^\dagger\frac{\ha_{\frac{1}{2}m}\hb_{-\frac{1}{2}m'}-\ha_{-\frac{1}{2}m}\hb_{\frac{1}{2}m'}}{\sqrt{2}} \notag\\&+ \rm{H.c}.
\end{align}
where $\vec{r}$ is the position of the atoms, and $\hbar\vec{k}$ and $\omega$ are the momentum and frequency of the laser photon inducing molecule formation. The detuning between the atomic and molecular states is $\hbar\omega-E_b$, and $\Omega$ is the transition matrix element from the atomic to the molecular state.

For large detuning, the occupation in the molecular state will be small. Therefore we can adiabatically eliminate the molecular state and obtain an effective interaction between the \Li\ and \Rb\ atoms using second-order perturbation theory:
\begin{equation}\begin{split}
 \hat{\tilde H}_{\rm{int}} = &\sum\limits_{mm'} \frac{\Omega^2}{E_b-\hbar\omega}\left(\frac{\ha_{\frac{1}{2}m}\hb_{-\frac{1}{2}m'}-\ha_{-\frac{1}{2}m}\hb_{\frac{1}{2}m'}}{\sqrt{2}}\right)\+ \\
 &\times\left(\frac{\ha_{\frac{1}{2}m}\hb_{-\frac{1}{2}m'}-\ha_{-\frac{1}{2}m}\hb_{\frac{1}{2}m'}}{\sqrt{2}}\right). \label{eqn:AC Stark shift}
\end{split}\end{equation}
Using Eq. (\ref{eqn:CGcoeffs}), the operators $\ha_{m_Sm_I}\+$ and $\hb_{m_Sm_I}\+$ can be projected into the hyperfine eigenstate basis. Assuming that the chemical potential is set such that the $F=3/2$ and $F=2$ manifolds are unoccupied, we project $\hat{\tilde H}_{\rm{int}}$ into the $F=1/2$ and $F=1$ manifolds. We obtain an effective interaction
\begin{equation}
 \hat{\tilde{\tilde H}}_{\rm{int}} = \frac{\Omega^2}{E_b-\hbar\omega} \sum_{\alpha\beta\mu\nu} \ha_\alpha\+\ha_\beta\hb_\mu\+\hb_\nu \left(- \frac{1}{12}\vec{\sigma}^{(a)}_{\alpha\beta}\cdot\vec{\sigma}^{(b)}_{\mu\nu} + \frac{1}{4}\delta_{\alpha\beta}\delta_{\mu\nu} \right).
 \label{eqn:effectiveInteraction}
\end{equation}

The first term in Eq. (\ref{eqn:effectiveInteraction}) is of the form of Kondo-like interactions with $g_s = \frac{-1}{12}\frac{\Omega^2}{E_b-\hbar\omega}$, and the second term a density-density interaction with $g_n = \frac{1}{4}\frac{\Omega^2}{E_b-\hbar\omega}$, where $g_s$ and $g_n$ were defined in Eq. (\ref{eqn:HKondoRealSpace}). Generally, in addition there would also be intrinsic interactions which modify the values of $g_s$ and $g_n$ in the experiment. To explore Kondo physics, $g_s$ should be positive.

If one wanted to exactly produce the Kondo model (where $g_n=0$), one could add more photoassociation lasers, for example, coupling the electronic spin-triplet atomic states. However as we show in Sec. \ref{sec:scattering}, the presence of a nonzero $g_n$ does not change the physics.

\subsection{Experimental and model parameters}
In this section we estimate our model parameters $g_s$ and $g_n$ for a typical experiment performing optical Feshbach resonance. We also discuss the issue of atom losses in optical Feshbach resonances.

Experiments implementing optical Feshbach resonances typically suffer from high atom loss rates because lasers bring the atomic states close to resonance with a bound molecular state. The excited molecular states have a finite linewidth, and either dissociate into free atoms with large kinetic energies or spontaneously decay to ground molecular states. The effect of a finite linewidth can be incorporated by making the ac Stark shift obtained in Eq. (\ref{eqn:AC Stark shift}) complex:
\begin{equation}
 g = \frac{\Omega^2}{E_b-\hbar\omega+i\hbar\gamma}.
\end{equation}
The real part of $g$, $Re(g) = \Omega^2\frac{E_b-\hbar\omega}{(E_b-\hbar\omega)^2+(\hbar\gamma)^2}$, is a measure of the interaction strength, and determines the magnitude of the model parameters $g_s$ and $g_n$. The magnitude of the imaginary part of $g$, $K_{\rm{PA}} = \Omega^2\frac{\hbar\gamma}{(E_b-\hbar\omega)^2+(\hbar\gamma)^2}$, is the inelastic collision rate co-efficient.

In experiments in which $^7$Li and $^{85}$Rb atoms are resonantly coupled to a molecular state, the inelastic collision rate co-efficient typically has a value $|K_{\rm{PA}}| \simeq \frac{\Omega^2}{\hbar\gamma} \sim 4\times10^{-11}\hbar\rm{cm}^3/\rm{s}$ for a moderate laser intensity of $100\rm{W}/\rm{cm}^2$ \cite{opticalFeshbachResonances2}. Typical linewidths are $\gamma\sim 10$MHz. We expect that $\gamma$ and $K_{\rm{PA}}$ would have similar values for any other alkali-metal combination, and in particular for \Li\ and \Rb\ as well. We note that $\Omega^2$ is proportional to the laser intensity. The inelastic collision rate can be reduced by increasing the detuning of the laser. If the laser detuning is 10 times the linewidth ($|E_b-\hbar\omega|=10\hbar\gamma$), then $K_{\rm{PA}} \sim  4\times 10^{-13}\hbar\rm{cm}^3/\rm{s}$ and $Re(g) \sim 4\times10^{-12}\hbar\rm{cm}^3/\rm{s}$ for a laser intensity of $100\rm{W}/\rm{cm}^2$.

The relevant quantities for estimating the temperature scale for observing Kondo physics are $\frac{g_sN}{V\epsilon_F}$ and $\frac{g_nN}{V\epsilon_F}$ where $\frac{N}{V}$ is the density of fermions. We find in Sec. \ref{sec:scattering} that for this minimum to occur at a temperature of $O(0.05T_F)$, $\frac{|g_{s,n}|N}{V\epsilon_F}$ should be $O(0.1)$. This can be achieved with a density of $\frac{N}{V}\sim10^{13}\rm{cm}^{-3}$ and an interaction strength $|g_{s,n}| \sim 2\times10^{-9} \hbar \rm{cm}^3/\rm{s}$, which requires roughly $500$ times larger intensity than that in \cite{opticalFeshbachResonances2}. A judicious choice of the resonance may significantly reduce the intensity required.

\section{Kondo-enhanced scattering between \Rb\ and \Li}\label{sec:scattering}
Here we calculate the momentum transfer in a collision between a fermionic cloud and a bosonic cloud. We show that spin-exchange collisions lead to a logarithmic temperature dependence of the momentum transferred. This logarithm is characteristic of the Kondo effect, and analogous to the behavior of electrical resistance of magnetic alloys. As shown in Fig. \ref{fig:temperature plot}, it leads to a minimum in the momentum transferred. The most naive way to measure this momentum exchanged would be to launch the Bose gas into a stationary Fermi gas and measure the final momentum of the Fermi gas. We briefly consider an alternative method in Sec. \ref{subsec:dipole modes}.

The duration of interaction between a boson and the Fermi gas in the experiment described above is $t=L/v$ where $L$ is the size of the Fermi cloud. We calculate the momentum transferred from the Bose gas to the Fermi gas at time $t$ to zeroth order in $1/M_b$, first order in $\vec{v}$, and third order in the interaction parameters $g_s$ and $g_n$. We perform this calculation for general values of $g_s$ and $g_n$ that are independent of each other. At the end of our calculation we specialize to the values of $g_s$ and $g_n$ produced by our proposal in Sec. \ref{sec:engineering interaction}. Since $L$ is a macroscopic quantity and we work in the small $v$ limit, we make a long time approximation wherever possible. We assume that the Bose gas is dilute, and neglect events involving scattering of a fermion with more than one boson. Equivalently we calculate the momentum transferred by one boson with momentum $M_b\vec{v}$, and sum over all bosons. The Fermi surface will play an important role.

We consider the collision of the Fermi gas with one boson with spin projection $m$ at time $0$. The momentum of the Fermi gas at time $t$ is then $\vec{P}_m = \frac{V}{(2\pi)^3}\int d^3\vec{k}\sum_\alpha\hbar\vec{k}n_{k\alpha m}(t)$, where
\begin{equation}
 n_{k\alpha m}(t) = \langle \hb_{M_bv,m}(0)\ha_{k\alpha}\+(t)\ha_{k\alpha}(t)\hb_{M_bv,m}\+(0)\rangle
 \label{eqn:n}
\end{equation}
is the occupation of fermions with momentum $\vec{k}$ and spin projection $\alpha$ at time $t$. In Eq. (\ref{eqn:n}) the expectation value is taken over a thermal ensemble of fermions, with no bosons present. The bosonic creation operator preceding the ket state in Eq. (\ref{eqn:n}) ensures that we calculate the occupation $n_k$ after the collision of \textit{one} boson with the Fermi gas. Since the bosons are spin unpolarized, the average momentum imparted by a boson is $\vec{P}_{av} = \frac{V}{3(2\pi)^3}\int d^3\vec{k}\sum_{\alpha m}\hbar\vec{k}n_{k\alpha m}(t)$. Multiplying by $N_b$, the number of bosons, the net momentum of the Fermi gas is
\begin{equation}
 \vec{P}(t) = \frac{N_bV}{3(2\pi)^3}\int d^3\vec{k}\sum_{\alpha m}\hbar\vec{k}n_{k\alpha m}(t).
 \label{eqn:final momentum definition}
\end{equation}

In appendix \ref{sec:calculation}, we describe our diagrammatic perturbation theory approach for calculating $n_{k\alpha m}(t)$.
We find that
%\begin{align}
% &\frac{1}{3}\sum_m n_{k\alpha m}(t) = f_k - \frac{4t\vec{k}\cdot\vec{v}\rho(\epsilon_k)}{V^2} \frac{\partial f_k}{\partial\epsilon_k}\label{eqn:nResult}\\
%& \times\left(\frac{S(S+1)}{4}\tilde{g}_s^2+\tilde{g}_n^2 - \frac{\tilde{g}_s^3S(S+1)}{4(2\pi)^3} \int d^3\vec{p}\frac{f_p}{\epsilon_k-\epsilon_p}\right),\notag
%\end{align}
\begin{align}
 \frac{1}{3}\sum_m n_{k\alpha m}(t) = &f_k - \frac{4t\vec{k}\cdot\vec{v}\rho(\epsilon_k)}{V^2} \frac{\partial f_k}{\partial\epsilon_k}\notag\\
& \times\left(\frac{S(S+1)}{4}\tilde{g}_s^2+\tilde{g}_n^2 - \frac{\tilde{g}_s^3S(S+1)}{4(2\pi)^3}\right.\notag \\
& \left.\times\int d^3\vec{p}\frac{f_p}{\epsilon_k-\epsilon_p}\right),\label{eqn:nResult}
\end{align}
plus terms which scale as $t^0, v^2$ or $1/M_b$. Due to our use of point interactions, the interaction parameters $g_s$ and $g_n$ are renormalized to $\tilde{g}_s$ and $\tilde{g}_n$. These renormalized (physical) coupling constants are the ones appearing in Eq. (\ref{eqn:nResult}). This renormalization of the interaction strength occurs at all orders of perturbation theory.

To calculate $\vec{P}(t)$, we sum the contributions due to all momentum states, and include the temperature dependence of the fermionic chemical potential, $\mu = \epsilon_F\left(1-\frac{\pi^2}{12}\left(\frac{k_BT}{\epsilon_F}\right)^2\right) + O\left(\frac{k_BT}{\epsilon_F}\right)^4$. We find that at long times,
%\begin{widetext}
%\begin{equation}\begin{array}{rcl}
% \vec{P} &=& \frac{3S(S+1)N_b}{8}\left(\frac{J}{\epsilon_F}\right)^2\left(k_FL\right)\hbar k_F \times
% \\ &&\left( \left(1+\frac{4}{S(S+1)}\left(\frac{\tilde{g}_n}{\tilde{g}_s}\right)^2\right)\left(1 + \frac{\pi^2}{6}\left(\frac{k_BT}{\epsilon_F}\right)^2\right) - \frac{3J}{2\epsilon_F} \left(1.13 +  \left(2.6-\frac{\pi^2}{48}\right)\left(\frac{k_BT}{\epsilon_F}\right)^2 + \frac{1}{2}\log\frac{k_BT}{4\epsilon_F}\left(1+\frac{5\pi^2}{12}\left(\frac{k_BT}{\epsilon_F}\right)^2\right) \right)\right),
%\end{array}\label{eqn:Kondo result}\end{equation}
%\end{widetext}
\begin{align}
 \vec{P} =& \frac{3S(S+1)N_b}{8}\left(\frac{J}{\epsilon_F}\right)^2\left(k_FL\right)\hbar k_F\notag\\
 & \times \left( \left(1+\frac{4}{S(S+1)}\left(\frac{\tilde{g}_n}{\tilde{g}_s}\right)^2\right)\left(1 + \frac{\pi^2}{6}\left(\frac{k_BT}{\epsilon_F}\right)^2\right)\right.\notag\\
 & - \frac{3J}{2\epsilon_F} \left(1.13 +  \left(2.6-\frac{\pi^2}{48}\right)\left(\frac{k_BT}{\epsilon_F}\right)^2 \right.\notag\\
 & \left.\left.+ \frac{1}{2}\log\frac{k_BT}{4\epsilon_F}\left(1+\frac{5\pi^2}{12}\left(\frac{k_BT}{\epsilon_F}\right)^2\right) \right)\right),
\label{eqn:Kondo result}
\end{align}
where $J = \tilde{g}_s\frac{N}{V}$, and $\frac{N}{V}$ is the density of fermions. In Eq. (\ref{eqn:Kondo result}) we have neglected terms which scale as $t^0,v^2,\frac{1}{M_b}$ or $T^4$. According to our proposal in Sec. \ref{sec:engineering interaction}, $\frac{\tilde{g}_n}{\tilde{g}_s}=-3$ and $S=1$. The result of Eq. (\ref{eqn:Kondo result}) is plotted in Fig. \ref{fig:temperature plot} using these parameters and $J=0.1\epsilon_F$. For comparison, we also plot the momentum transferred to the Fermi gas for spin-independent interactions with the same value of $\tilde{g}_n = -3\times\frac{0.1\epsilon_F}{N/V}$ and $\tilde{g}_s=0$. The logarithmic temperature dependence of $\vec{P}$ for spin-dependent interactions is characteristic of Kondo physics. Equation (\ref{eqn:Kondo result}) breaks down when $\frac{J}{\epsilon_F}\log\frac{k_BT}{\epsilon_F} \simeq O(1)$. Below this temperature, the logarithmic increase saturates to a constant. Calculation of this saturation is the subject of the Kondo problem and can be addressed with renormalization group or Bethe ansatz methods. Equation (\ref{eqn:Kondo result}) also breaks down when $v\simeq \frac{k_BT}{\hbar k_F}$.

The momentum transferred $|\vec{P}|$ has a minimum at a temperature $T_{\rm{min}} \sim \frac{3}{2\pi k_B}\sqrt{\frac{J\epsilon_F}{1+\frac{4\tilde{g}_n^2}{\tilde{g}_s^2S(S+1)}}}$. For the parameters $\frac{\tilde{g}_n}{\tilde{g}_s} = -3$, $S=1$, and $J=0.1\epsilon_F$, this minimum occurs at a temperature $\frac{T}{T_F}\simeq O(0.05)$. At this temperature and interaction strength, the momentum imparted by one boson to the Fermi gas is $\frac{|\vec{P}|}{N_b} \simeq \frac{3}{4}\hbar k_F\left(\frac{J}{\epsilon_F}\right)^2\left(k_FL\right)$. For a $20\mu$m long Fermi cloud at a density of $10^{13}\rm{cm}^{-3}$, the momentum imparted per boson is nearly $1.2\hbar k_F$. We estimated in Sec. \ref{sec:engineering interaction} that achieving $J=0.1\epsilon_F$ would require high intensity lasers and tight trapping of the fermions. The observation of this minimum will be a direct experimental confirmation of Kondo physics.

\subsection{Alternative methods to measure enhanced Kondo scattering} \label{subsec:dipole modes}
Here we briefly explain an alternative method to measure the enhanced Kondo scattering between a Fermi cloud and a Bose cloud. We consider inducing dipole oscillations of a Bose cloud and a Fermi cloud in a harmonic trap of frequency $\omega$. The clouds will collide every half-cycle and exchange momentum $\vec{P}$. As a result the amplitude of oscillations of the Fermi cloud will reduce each half cycle. Conservation of momentum implies that the maximum fermion displacement $X$ will reduce each half cycle by $\delta X \simeq \frac{|\vec{P}|}{N_am_a\omega}$ where $N_a$ is the number of fermions; the Bose cloud's amplitude will not change very much because of the bosons' heavy mass. The Bose-Fermi interaction interval is longer for a smaller relative momentum, and vice versa. Thus the momentum exchanged $|\vec{P}|$ is independent of the relative velocities of the cloud, leading to a linear decay of the amplitude rather than exponential; $\frac{d\delta X}{dt} \sim \frac{|\vec{P}|}{N_am_a\pi}$. If the Bose-Fermi interactions are Kondo-like, the damping rate of amplitude of oscillations will have a minimum at the same temperature as $|\vec{P}|$ does, $T_{\rm{min}} \sim \frac{3}{2\pi k_B}\sqrt{\frac{J\epsilon_F}{1+\frac{4\tilde{g}_n^2}{\tilde{g}_s^2S(S+1)}}}$. For a typical amplitude of oscillation $X\simeq 100\mu$m in a trap of frequency $\omega=2\pi\times 10$ Hz, and if $\frac{N_a}{N_b}=200$, the amplitude will decay to zero in about 12 oscillations at $T=T_{\rm{min}}$. The observation of a minimum in the damping rate will also be an experimental confirmation of Kondo physics.

\section{Summary}\label{sec:summary}
We considered scattering between a spin-1/2 Fermi gas and a dilute spin-unpolarized Bose gas. As an example we considered \Li\ and \Rb\ as our itinerant fermions and bosonic magnetic impurities. We proposed using an optical Feshbach resonance to produce rotationally symmetric interactions between the \Li\ and \Rb\ atoms, which included both spin-dependent Kondo-like and spin-independent density-density interactions. We argued that these interactions would give rise to enhanced Fermi-Bose scattering. We perturbatively calculated the temperature dependence of the momentum transferred to the Fermi gas in a scattering experiment, up to third order in the Bose-Fermi interaction strength. We showed that the temperature dependence of the momentum transferred has a minimum at a characteristic temperature and is logarithmic at low temperatures, characteristic of the Kondo effect and analogous to the behavior of electrical resistance in magnetic alloys.

Our proposal to implement spin-dependent interactions requires overcoming significant experimental challenges such as using high intensity lasers to achieve large interaction strengths. However, overcoming these challenges enable the possibility of exploring exotic phenomena due to Kondo physics. The ground state of a Bose-Fermi mixture with Kondo-type spin-dependent interactions should display interesting correlations, with each boson surrounded by a screening cloud of fermions with opposite spin \cite{affleckKondoScreeningReview}. These clouds may be observable through various imaging techniques \cite{quantumMicroscopeZwierlein,quantumMicroscopeGreiner,quantumMicroscopeMiranda}. Similar experiments with bosons confined to a lattice would probe an analog of the Kondo lattice problem.

One can explore other techniques to experimentally produce Kondo-type interactions. For example, optically coupling the electronic triplet states of \Li-\Rb\ with excited molecular states will lead to a rotationally asymmetric interaction which also displays Kondo physics. Alternatively, one can realize the Anderson model and Kondo-like situations by trapping impurities in deep potentials \cite{AMReyAnderson,demlerAnderson,yusukeAnderson,anotherAnderson}.

\section*{ACKNOWLEDGEMENTS}
We acknowledge support from ARO-MURI Non-equilibrium Many-body Dynamics grant (Grant no. W911NF-14-1-0003). We would like to thank Kirk Madison for useful discussions including suggesting the damping of dipole modes as a probe. We would like to thank Todd Rutkowski for independently verifying the calculations in Sec. \ref{sec:engineering interaction}.

\appendix
\section{CALCULATION OF THE MOMENTUM TRANSFERRED}\label{sec:calculation}
\begin{figure}[t]
 \begin{center}
 \includegraphics[width=0.9\columnwidth,height=1.6cm]{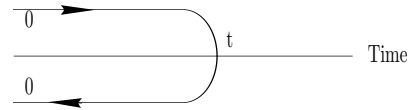}
 \end{center}
\caption{In our integrals, time begins at $0$, passes through $t$, then returns to $0$. Our perturbation theory requires ordering operators along this path.}
\label{fig:path}
\end{figure}

Here we calculate $n_{k\alpha m}(t)$ in Eq. (\ref{eqn:n}) and $\vec{P}(t)$ in Eq. (\ref{eqn:final momentum definition}). The standard way to calculate quantities like $n_{k\alpha m}(t)$ is using the $S$-matrix \cite{mahan}:
\begin{equation}\begin{split}
 n_{k\alpha m}(t) &= \langle T \hat{S}\hb_{M_bv,m}(0)\ha_{k\alpha}\+(t)\ha_{k\alpha}(t)\hb_{M_bv,m}\+(0)\rangle_0,\\
 \hat{S} &= e^{-i\int d\tau \HInt(\tau)},
\end{split}\label{eqn:path ordering}\end{equation}
where $T$ orders the operators along a path shown in Fig. \ref{fig:path} which starts at time $0$, passes through time $t$, and returns to time $0$.
All our integrals over time follow this path. The notation $\langle\rangle_0$ implies that all operators inside $\langle\rangle_0$ evolve according to
\begin{equation}\begin{split}
 \ha_{k\alpha}(t) &= e^{i\hH_0t/\hbar}\ha_{k\alpha}e^{-i\hH_0t/\hbar},\\
 \hb_{k\mu}(t) &= e^{i\hH_0t/\hbar}\hb_{k\mu}e^{-i\hH_0t/\hbar},
\end{split}\end{equation}
and states are weighted by $e^{-\beta\hH_0}$. Since $\hH_0$ is quadratic in $\ha_{k\alpha}$ and $\hb_{k\mu}$, the right-hand side of $n_{k\alpha m}(t)$ in Eq. (\ref{eqn:path ordering}) can be contracted using Wick's theorem. As a result, $n_{k\alpha m}(t)$ can be expressed diagrammatically as a sum of Feynman's diagrams. We calculate these Feynman's diagrams up to $O(g_s^3)$ and $O(g_n^3)$ in the long time limit.

\subsection{Feynman rules}
\begin{figure}[t]
\includegraphics[width=0.9\columnwidth]{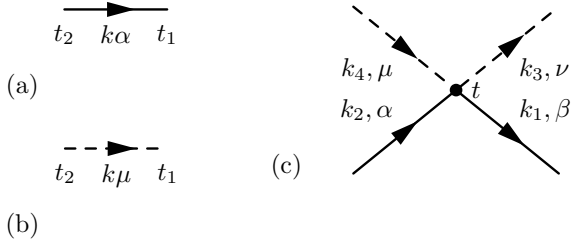}
\caption{Diagrammatic representation of vertex and propagators. (a) Solid line denotes a fermion propagator which propagates a fermion with momentum $k$ and spin projection $\alpha$ from time $t_2$ to $t_1$. (b) Dashed line denotes a boson propagator which propagates a boson with momentum $k$ and spin projection $\mu$ from time $t_2$ to $t_1$. (c) A vertex denotes the matrix element for a Bose-Fermi scattering event. Mathematical expressions are given in Eqs.(\ref{eqn:Feynman rules for propagators}) and (\ref{eqn:Feynman rule for vertex}).}
\label{fig:propagators and vertex}
\end{figure}

We denote the propagator for fermions, $\langle T \ha_{k\alpha}(t_1)\ha_{k\alpha}\+(t_2)\rangle_0$, by a solid line, and the propagator for bosons, $\langle T \hb_{k\mu}(t_1)\hb_{k\mu}\+(t_2)\rangle_0$, by a dotted line, depicted in Figs. \ref{fig:propagators and vertex}(a) and \ref{fig:propagators and vertex}(b). Their values are
\begin{equation}\begin{split}
 \langle T \ha_{k\alpha}(t_1)\ha_{k\alpha}\+(t_2)\rangle_0 &= e^{-i\epsilon_k(t_1-t_2)}\left(\Theta(t_1-t_2)-f_k\right),\\
 \langle T \hb_{k\mu}(t_1)\hb_{k\mu}\+(t_2)\rangle_0 &= e^{-iE_k(t_1-t_2)}\Theta(t_1-t_2).
\end{split}\label{eqn:Feynman rules for propagators}\end{equation}
In Eq. (\ref{eqn:Feynman rules for propagators}), $\Theta(t_1-t_2)=1$ if $t_1$ is after $t_2$ along the path in Fig. \ref{fig:path}, and 0 otherwise.

We perturbatively expand $n_{k\alpha m}(t)$ in the vertex depicted in Fig. \ref{fig:propagators and vertex}(c), whose value is
\begin{equation}\begin{split}
\raisebox{-1cm}{\includegraphics{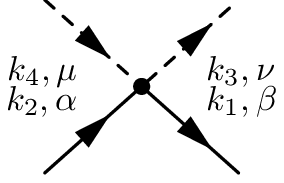}} \ = &\frac{(2\pi)^3}{V^2}\delta(k_1+k_3-k_2-k_4)\\
  & \times\left(g_s\vec{\sigma}^{(1/2)}_{\alpha\beta}\cdot\vec{\sigma}^{(S)}_{\mu\nu} + g_n\delta_{\alpha\beta}\delta_{\mu\nu}\right).
\label{eqn:Feynman rule for vertex}
\end{split}\end{equation}

The vertex denotes a scattering event between a fermion and a boson. The time at which this scattering event occurs is integrated over the path in Fig. \ref{fig:path}. All momenta and spin projections are summed/integrated over, with the constraint that momenta and spin are conserved at each vertex. The diagrams which contribute to Eq. (\ref{eqn:path ordering}) have four external propagators. There is an incoming and outgoing fermion propagator evaluated at time $t$, and carrying momentum $\hbar\vec{k}$ and spin projection $\alpha$. There is also an incoming and outgoing boson propagator evaluated at time $0$, and carrying momentum $M_b\vec{v}$ and spin projection $m$. All lines and vertices in a Feynman diagram can be labeled using the rules described above. Therefore we omit labels. Finally, each diagram carries a multiplicity, which is the number of times it appears in the expansion of Eq. (\ref{eqn:path ordering}) in powers of $g_s$ and $g_n$.

\subsection{Calculation of $n_{k\alpha m}(t)$}
Terms of $O(g_{s,n}^n)$ in the perturbative expansion of $n_{k\alpha m}(t)$ contain $2n+2$ pairs of operators leading to $(n+1)!^2$ contractions. The resulting number of diagrams increases exponentially with $n$. We explicitly consider each order and evaluate the nonzero diagrams.

\subsubsection{Zeroth order}
\begin{figure}[t]
 \unitlength=1in
 \begin{center}
 \includegraphics[width=0.3\columnwidth]{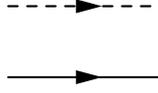}
 \end{center}
  \caption{Zeroth-order diagram in the expansion for $n_{k\alpha m}(t)$.}
  \label{fig:zeroth order}
\end{figure}

The expression for the zeroth-order term in the expansion of $n_{k\alpha m}(t)$ is
\begin{equation}
 n_{k\alpha m}^{(0)}(t) = \langle T \hb_{M_bv,m}(0)\ha_{k\alpha}\+(t)\ha_{k\alpha}(t)\hb_{M_bv,m}\+(0)\rangle_0.
\end{equation}
Using Wick's theorem,
\begin{equation}\begin{split}
 n_{k\alpha m}^{(0)}(t) &= \langle \hb_{M_bv,m}(0)\hb_{M_bv,m}\+(0)\rangle_0 \langle\ha_{k\alpha}\+(t)\ha_{k\alpha}(t)\rangle_0\\
 &= f_k.
\end{split}\end{equation}
The corresponding Feynman diagram is shown in Fig. \ref{fig:zeroth order}. Since the bosons and fermions do not interact at this order, $n_{k\alpha m}^{(0)}$ does not contribute to any momentum transfer.

\subsubsection{First order}\label{subsubsec:one}
The first-order term in the expansion for $n_{k\alpha m}(t)$ is
\begin{equation}\begin{split}
 n_{k\alpha m}^{(1)}(t) =& -i\int d\tau_1 \\ &\langle T \HInt(\tau_1)\hb_{M_bv,m}(0)\ha_{k\alpha}\+(t)\ha_{k\alpha}(t)\hb_{M_bv,m}\+(0)\rangle_0.
\end{split}\end{equation}
\begin{figure}[t]
\includegraphics[width=1.0\columnwidth]{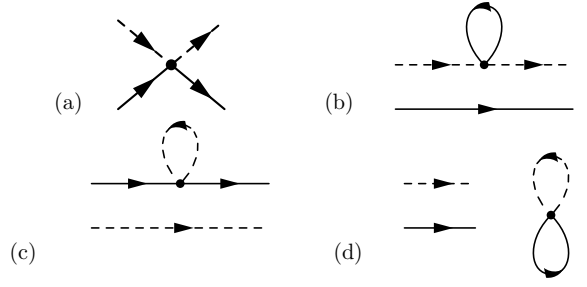}
\caption{First-order diagrams in the expansion of $n_{k\alpha m}(t)$.}
\label{fig:first order}
\end{figure}
By Wick-contracting the above expression, we find that $n_{k\alpha m}^{(1)}(t)$ is the sum of the four diagrams shown in Fig. \ref{fig:first order}, all of which evaluate to zero. For example,
\begin{equation}
\raisebox{-0.7cm}{\includegraphics{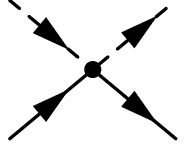}}\ = \int d\tau \left(\frac{1}{2}g_sm + g_n\right) = 0.
\end{equation}
Due to the same reason, Figs. \ref{fig:first order}(b), \ref{fig:first order}(c) and \ref{fig:first order}(d) are also zero. Therefore,
\begin{equation}
 \frac{1}{3}\sum_m n_{k\alpha m}^{(1)}(t) = 0.
\end{equation}
Moreover, the same reasoning implies that all higher-order diagrams in which a fermion or boson loop begins and ends at the same vertex are also zero.

\subsubsection{Second order}\label{subsubsec:two}

\begin{figure}
\includegraphics[width=1.4\columnwidth]{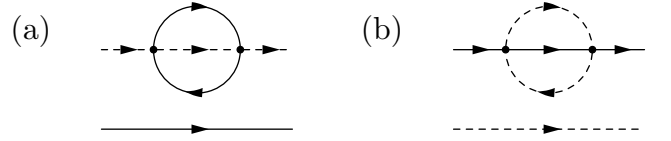}
\caption{Two of the diagrams that are zero at second order.}
\label{fig:mass corrections second order}
\end{figure}
\begin{figure}
\includegraphics[width=0.9\columnwidth]{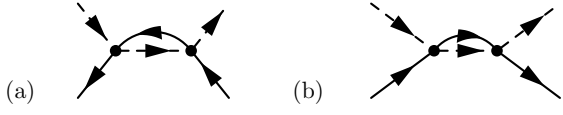}
\caption{Nonzero diagrams at $O(g^2)$ in the expansion for $n_{k\alpha m}(t)$.}
\label{fig:second order}
\end{figure}

The second-order term,
\begin{align}
 n_{k\alpha m}^{(2)}(t) = &-\frac{1}{2}\int d\tau_1d\tau_2 \langle T \HInt(\tau_1)\HInt(\tau_2)\hb_{M_bv,m}(0)\notag\\
 &\times\ha_{k\alpha}\+(t)\ha_{k\alpha}(t)\hb_{M_bv,m}\+(0)\rangle_0,
\end{align}
can be contracted into Wick pairs in 36 ways, which give rise to 20 different diagrams. Most of these diagrams are zero because of reasons explained in Sec. \ref{subsubsec:one}. In addition, the diagrams shown in Fig. \ref{fig:mass corrections second order} also evaluate to zero. For example, since we work in the dilute boson limit, there can only be one boson line in any time slice, implying that Fig. \ref{fig:mass corrections second order}(a) is zero. The only two nonzero diagrams are shown in Fig. \ref{fig:second order}.

Using our Feynman rules,
%\begin{widetext}
\begin{align}
\frac{1}{3}\sum_m\raisebox{-0.2cm}{\includegraphics[width=0.17\columnwidth]{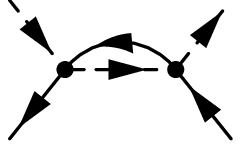}} = &\frac{2}{V(2\pi)^3}\int d^3\vec{p}\ (1-f_k)f_p\frac{\sin^2\delta\epsilon t/\hbar}{\delta\epsilon^2}\notag\\
& \times\left(g_s^2\frac{S(S+1)}{2}+2g_n^2\right),
\end{align}
%\begin{equation}
% \frac{1}{3}\sum_m\raisebox{-0.2cm}{\includegraphics[width=0.12\columnwidth]{figA11.eps}} = \frac{2}{V(2\pi)^3}\int d^3\vec{p}\ (1-f_k)f_p\frac{\sin^2\delta\epsilon t/\hbar}{\delta\epsilon^2} \left(g_s^2\frac{S(S+1)}{2}+2g_n^2\right),
%\end{equation}
and
\begin{align}
\frac{1}{3}\sum_m \raisebox{-0.2cm}{\includegraphics[width=0.2\columnwidth]{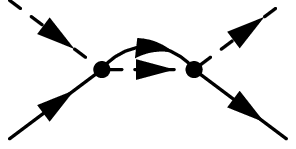}} = &-\frac{2}{V(2\pi)^3}\int d^3\vec{p}\ f_k(1-f_p)\frac{\sin^2\delta\epsilon' t/\hbar}{\delta\epsilon'^2}\notag\\
& \times\left(g_s^2\frac{S(S+1)}{2}+2g_n^2\right),
\end{align}
%\begin{equation}
% \frac{1}{3}\sum_m \raisebox{-0.2cm}{\includegraphics[width=0.12\columnwidth]{figA12.eps}} = -\frac{2}{V(2\pi)^3}\int d^3\vec{p}\ f_k(1-f_p)\frac{\sin^2\delta\epsilon' t/\hbar}{\delta\epsilon'^2} \left(g_s^2\frac{S(S+1)}{2}+2g_n^2\right),
%\end{equation}
where $\delta\epsilon = \frac{1}{2}\left(\epsilon_k-\epsilon_p-\frac{1}{2}M_bv^2+\frac{(\hbar\vec{k}-\hbar\vec{p}-M_b\vec{v})^2}{2M_b}\right)$ and $\delta\epsilon' = \frac{1}{2}\left(\epsilon_k-\epsilon_p+\frac{1}{2}M_bv^2-\frac{(\hbar\vec{k}-\hbar\vec{p}+M_b\vec{v})^2}{2M_b}\right)$. Neglecting terms of order $1/M_b$, $\delta\epsilon=\delta\epsilon'=\frac{1}{2}\left(\epsilon_{k-m_av/\hbar}-\epsilon_{p-m_av/\hbar}\right)$. The resulting second-order contribution is
\begin{equation}%\begin{split}
 \frac{1}{3}\sum_mn_{k\alpha m}^{(2)}(t) = -\frac{g_s^2\frac{S(S+1)}{2} +2g_n^2}{V(2\pi)^3}
  \int d^3\vec{p}\ (f_k-f_p)\frac{\sin^2t\delta\epsilon/\hbar}{\delta\epsilon^2}.
%\end{split}
\end{equation}
Since the bosons are much heavier than the fermions, they have nearly the same velocity $\vec{v}$ before and after scattering. Therefore, it is easier to work in the bosons' rest frame. For small $\vec{v}$,
\begin{equation}\begin{split}
 \frac{1}{3} \sum_m n_{k+\frac{m_av}{\hbar},\alpha m}^{(2)}(t)=& -\frac{g_s^2\frac{S(S+1)}{2}+2g_n^2}{V(2\pi)^3}\int d^3\vec{p} \\&\times
  \left(f_k-f_p + \hbar\vec{k}\cdot\vec{v}\frac{\partial f_k}{\partial\epsilon_k}-\hbar\vec{p}\cdot\vec{v}\frac{\partial f_p}{\partial\epsilon_p}\right)\\ &\times
  \frac{\sin^2(t(\epsilon_k-\epsilon_p)/2\hbar)}{((\epsilon_k-\epsilon_p)/2)^2} + O(v^2,1/M_b)
\end{split}
\label{eqn:two}\end{equation}
where $O(v^2,1/M_b)$ refers to terms which scale as $v^2$ or $1/M_b$. The first two terms in Eq. (\ref{eqn:two}) have negligible contribution near $\epsilon_k=\epsilon_p$. At long times, any significant contribution comes from the tail of $\frac{\sin^2(t(\epsilon_k-\epsilon_p)/2\hbar)}{((\epsilon_k-\epsilon_p)/2)^2}$, where $\sin^2(t(\epsilon_k-\epsilon_p)/2\hbar)$ can be approximated by its average, $1/2$. Hence their contribution saturates to a constant at long times. For the last two terms in Eq. (\ref{eqn:two}), which are significant near $\epsilon_k=\epsilon_p$, we approximate $\frac{\sin^2(t(\epsilon_k-\epsilon_p)/2\hbar)}{((\epsilon_k-\epsilon_p)/2)^2} \simeq\frac{2t\delta(\epsilon_k-\epsilon_p)}{\hbar}$. Hence at long times,
\begin{align}
 &\frac{1}{3}\sum_m  n_{k+\frac{m_av}{\hbar},\alpha m}^{(2)}(t) = -\frac{2(g_s^2\frac{S(S+1)}{2}+2g_n^2)t}{V(2\pi)^3}\\
 & \int d^3\vec{p} \left(\vec{k}\cdot\vec{v}\frac{\partial f_k}{\partial\epsilon_k}-\vec{p}\cdot\vec{v}\frac{\partial f_p}{\partial\epsilon_p}\right) \delta(\epsilon_k-\epsilon_p) + O\left(t^0,v^2,\frac{1}{M_b}\right)\notag\\
 &= -4\frac{g_s^2\frac{S(S+1)}{4}+g_n^2}{V^2} t\vec{k}\cdot\vec{v}\frac{\partial f_k}{\partial\epsilon_k}\rho(\epsilon_k) + O\left(t^0,v^2,\frac{1}{M_b}\right),\notag
\end{align}
%\begin{equation}\begin{split}
%\frac{1}{3}\sum_m  n_{k+\frac{m_av}{\hbar},\alpha m}^{(2)}(t) &= -\frac{2(g_s^2\frac{S(S+1)}{2}+2g_n^2)t}{V(2\pi)^3}
%  \int d^3\vec{p} \left(\vec{k}\cdot\vec{v}\frac{\partial f_k}{\partial\epsilon_k}-\vec{p}\cdot\vec{v}\frac{\partial f_p}{\partial\epsilon_p}\right) \delta(\epsilon_k-\epsilon_p) + O\left(t^0,v^2,\frac{1}{M_b}\right)\\
% &= -4\frac{g_s^2\frac{S(S+1)}{4}+g_n^2}{V^2} t\vec{k}\cdot\vec{v}\frac{\partial f_k}{\partial\epsilon_k}\rho(\epsilon_k) + O\left(t^0,v^2,\frac{1}{M_b}\right),
%\end{split}\end{equation}
where $\rho(\epsilon_k)$ is the three-dimensional density of states for a single spin projection. In the laboratory frame,
\begin{align}
 \frac{1}{3}\sum_m n_{k\alpha m}^{(2)}(t) =& -4\frac{g_s^2\frac{S(S+1)}{4}+g_n^2}{V^2} t\vec{k}\cdot\vec{v}\frac{\partial f_k}{\partial\epsilon_k}\rho(\epsilon_k)\notag
  \\& + O\left(t^0,v^2,\frac{1}{M_b}\right).
\end{align}
%\begin{equation}
%  \frac{1}{3}\sum_m n_{k\alpha m}^{(2)}(t) = -4\frac{g_s^2\frac{S(S+1)}{4}+g_n^2}{V^2} t\vec{k}\cdot\vec{v}\frac{\partial f_k}{\partial\epsilon_k}\rho(\epsilon_k) + O\left(t^0,v^2,\frac{1}{M_b}\right).
%\end{equation}
%\end{widetext}

\subsubsection{Third order}
\begin{figure}[t]
\includegraphics[width=1.5\columnwidth]{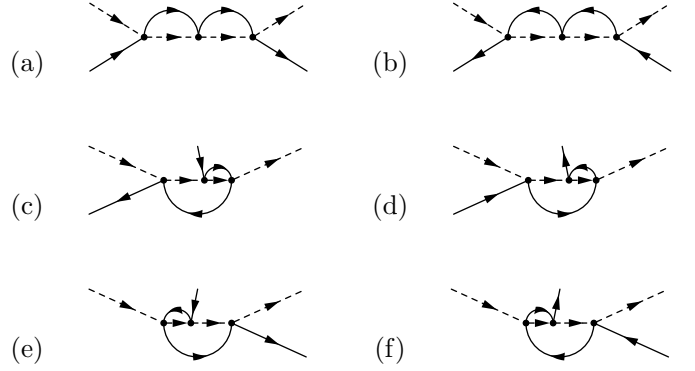}
\caption{Non-zero diagrams at $O(g^3)$ in the expansion for $n_{k\alpha m}(t)$.}
\label{fig:third order}
\end{figure}

The third-order term
\begin{align}
 &n_{k\alpha m}^{(3)}(t) = \frac{i}{6}\int d\tau_1d\tau_2 d\tau_3 \langle T \HInt(\tau_1)\times\\
 &\HInt(\tau_2)\HInt(\tau_3)\hb_{M_bv,m}(0)\ha_{k\alpha}\+(t)\ha_{k\alpha}(t)\hb_{M_bv,m}\+(0)\rangle_0\notag
\end{align}
can be contracted into Wick pairs in 576 ways. However due to reasons explained in Secs. \ref{subsubsec:one} and \ref{subsubsec:two} all diagrams except the ones shown in Fig. \ref{fig:third order} are zero. After a treatment similar to the one at second order, we calculate the third-order contribution to be
\begin{equation}\begin{split}
 \frac{1}{3}\sum_m n_{k\alpha m}^{(3)}(t) =& \frac{1}{V^2(2\pi)^3}\int d^3\vec{p}\ t\vec{v}\cdot\vec{k}\rho(\epsilon_k)\frac{\partial f_k}{\partial\epsilon_k} \frac{1}{\epsilon_k-\epsilon_p}\\
 & \times \left(f_pg_s^3S(S+1) - g_s^3\frac{S(S+1)}{2}\right.\\
 &\left. - 3g_s^2g_nS(S+1) - 4g_n^3\right) + O\left(t^0,v^2,\frac{1}{M_b}\right).\label{eqn:interesting result third order}
\end{split}\end{equation}
The right hand side of Eq. (\ref{eqn:interesting result third order}) consists of an ultraviolet divergent term arising from $\int d^3\vec{p} \frac{1}{\epsilon_k-\epsilon_p}$, and a finite term $\int d^3\vec{p}\frac{f_p}{\epsilon_k-\epsilon_p}$ which will ultimately give rise to a logarithmic temperature dependence. The ultraviolet divergence is an artefact of choosing a contact potential between the fermions and bosons which is nonzero only when they are at the same location in space. In reality, the interaction between the fermions and bosons has a finite range, which removes the ultraviolet divergence by introducing an upper cutoff on the limits on the integral over momenta. The exact details are unimportant if we express our results in terms of physical quantities. To this effect, we define effective coupling constants $\tilde{g}_s$ and $\tilde{g}_n$ where
\begin{equation}\begin{split}
 \tilde{g}_s^2 &= g_s^2\left(1 + \frac{g_s+6g_n}{2(2\pi)^3}\int d^3\vec{p} \frac{1}{\epsilon_k-\epsilon_p}\right),\\
 \tilde{g}_n^2 &= g_n^2\left(1 + \frac{g_n}{(2\pi)^3}\int d^3\vec{p} \frac{1}{\epsilon_k-\epsilon_p}\right).
\end{split}\end{equation}
The result for $n_{k\alpha m}(t)$ has no ultraviolet divergences when expressed in terms of $\tilde{g}_s$ and $\tilde{g}_n$.

The resulting $n_{k\alpha m}(t)$ at long times is
\begin{equation}\begin{split}
% \frac{1}{3}\sum_m n_{k\alpha m}(t) =& f_k - \frac{4t\vec{k}\cdot\vec{v}\rho(\epsilon_k)}{V^2} \frac{\partial f_k}{\partial\epsilon_k}\\
%& \times\left(\frac{S(S+1)}{4}\tilde{g}_s^2+\tilde{g}_n^2 - \frac{\tilde{g}_s^3S(S+1)}{4(2\pi)^3}\right.\\
% &\times \left.\int d^3\vec{p}\frac{f_p}{\epsilon_k-\epsilon_p}\right).
 &\frac{1}{3}\sum_m n_{k\alpha m}(t) = f_k - \frac{4t\vec{k}\cdot\vec{v}\rho(\epsilon_k)}{V^2} \frac{\partial f_k}{\partial\epsilon_k}\\
& \times\left(\frac{S(S+1)}{4}\tilde{g}_s^2+\tilde{g}_n^2 - \frac{\tilde{g}_s^3S(S+1)}{4(2\pi)^3}
 \int d^3\vec{p}\frac{f_p}{\epsilon_k-\epsilon_p}\right).
\end{split}\end{equation}
\\ \\
\subsection{Final momentum of the Fermi gas}
The total momentum $\vec{P}$ of the Fermi gas [defined in Eq. (\ref{eqn:final momentum definition})] will be along the direction of $\vec{v}$. Its magnitude is
\begin{equation}\begin{split}
 |\vec{P}| = &\frac{\vec{v}\cdot\vec{P}}{v} = -\frac{8t\hbar N_b }{vV(2\pi)^3}\int d^3\vec{k} \left(\vec{k}\cdot\vec{v}\right)^2 \frac{\partial f_k}{\partial\epsilon_k}\rho(\epsilon_k)
 \\&\times\left(\frac{S(S+1)}{4}\tilde{g}_s^2+\tilde{g}_n^2 - \frac{\tilde{g}_s^3S(S+1)}{4(2\pi)^3}\right.\\
 &\times \left. \int d^3\vec{p}\frac{f_p}{\epsilon_k-\epsilon_p}\right).
\end{split}\end{equation}
After integrating out the angular co-ordinates of $\vec{k}$ and $\vec{p}$ and performing a change of variables,
\begin{equation}\begin{split}
 |\vec{P}| = &-\frac{16m_aLN_b}{3\hbar V^2}\int d\epsilon\ \epsilon \frac{\partial f(\epsilon)}{\partial\epsilon}\rho^2(\epsilon)\\
 & \times\left(\frac{S(S+1)}{4}\tilde{g}_s^2+\tilde{g}_n^2 - \frac{\tilde{g}_s^3S(S+1)}{4V} \right.\\
 &\times\left. \int d\epsilon_p\frac{\rho(\epsilon_p)f(\epsilon_p)}{\epsilon-\epsilon_p}\right).
\end{split}\label{eqn:final momentum calculation}\end{equation}

We evaluate the second-order terms using a Sommerfield expansion,
\begin{equation}\begin{split}
|\vec{P}_2| \simeq &\frac{3m_aLN_b}{4\hbar\epsilon_F}J^2S(S+1)(1+\alpha^2)\left(1+\frac{\pi^2}{6}\left(\frac{k_BT}{\epsilon_F}\right)^2\right)
\\& + O\left(\frac{k_BT}{\epsilon_F}\right)^4.
\end{split}\end{equation}
where $J = \tilde{g}_s\frac{N}{V}$ and $\alpha = \frac{\tilde{g}_n}{\tilde{g}_s}\frac{2}{\sqrt{S(S+1)}}$.

\begin{widetext}
The third-order terms are
\begin{equation}\begin{split}
|\vec{P}_3| &= \frac{4S(S+1)m_aLN_b}{3\hbar}\left(\frac{\tilde{g}_s}{V}\right)^3 \int d\epsilon\ \epsilon \frac{\partial f(\epsilon)}{\partial\epsilon}\rho^2(\epsilon) \times \int d\epsilon_p\frac{\rho(\epsilon_p)f(\epsilon_p)}{\epsilon-\epsilon_p}\\
 &= -\frac{9S(S+1)m_aLN_b}{16\hbar\epsilon_F^{9/2}}J^3 \int_0^\infty d\epsilon\ \epsilon^2 \frac{\partial f(\epsilon)}{\partial\epsilon} \int_0^\infty d\epsilon_p \sqrt{\epsilon_p}\frac{f(\epsilon_p)}{\epsilon-\epsilon_p}.
\end{split}\end{equation}
We simplify the above expression by performing integration by parts,
\begin{equation}\begin{split}
|\vec{P}_3| &= \frac{9S(S+1)m_aLN_b}{8\hbar\epsilon_F^{9/2}}J^3 \int_0^\infty d\epsilon\ \epsilon^2 \frac{\partial f(\epsilon)}{\partial\epsilon}  \times\int_0^\infty d\epsilon_p f(\epsilon_p)\frac{\partial}{\partial\epsilon_p}\left(\sqrt{\epsilon_p}+\frac{\sqrt{\epsilon}}{2} \log\left|\frac{\sqrt{\epsilon}-\sqrt{\epsilon_p}}{\sqrt{\epsilon}+\sqrt{\epsilon_p}}\right|\right)\\
 &= -\frac{9S(S+1)m_aLN_b}{8\hbar\epsilon_F^{9/2}}J^3 \int_0^\infty d\epsilon\ \epsilon^2 \frac{\partial f(\epsilon)}{\partial\epsilon} \times\int_0^\infty d\epsilon_p \frac{\partial f(\epsilon_p)}{\partial\epsilon_p}\left(\sqrt{\epsilon_p}+\frac{\sqrt{\epsilon}}{2} \log\left|\frac{\beta(\epsilon-\epsilon_p)}{\beta(\sqrt{\epsilon}+\sqrt{\epsilon_p})^2}\right|\right).
\end{split}\label{eqn:third order}\end{equation}
We split Eq. (\ref{eqn:third order}) into two terms. We evaluate one of these terms numerically,
\begin{equation}
 \int_0^\infty d\epsilon\ \epsilon^{5/2} \frac{\partial f(\epsilon)}{\partial\epsilon} \int_0^\infty d\epsilon_p \frac{\partial f(\epsilon_p)}{\partial\epsilon_p}\log(\beta(\epsilon-\epsilon_p)) \simeq\epsilon_F^{5/2}\left(0.26 + 5.2\left(\frac{k_BT}{\epsilon_F}\right)^2\right) + O\left(\frac{k_BT}{\epsilon_F}\right)^4.
\end{equation}
We use a Sommerfield expansion for the remaining term. The result is
\begin{equation}
 |\vec{P}_3| \simeq  -\frac{9S(S+1)m_aLN_b}{8\hbar\epsilon_F^2}J^3
 \left(1.13+\left(2.6-\frac{\pi^2}{48}\right)\left(\frac{k_BT}{\epsilon_F}\right)^2+\frac{1}{2}\log\frac{k_BT}{4\epsilon_F}\left(1+\frac{5\pi^2}{12}\left(\frac{k_BT}{\epsilon_F}\right)^2\right)\right) + O\left(\frac{k_BT}{\epsilon_F}\right)^4.
\end{equation}

The final momentum of the Fermi gas is
\begin{equation}\begin{split}
 \vec{P} =& P_0\hat{v} \left( 1 + \frac{\pi^2}{6}\left(\frac{k_BT}{\epsilon_F}\right)^2 - \frac{3J}{2(1+\alpha^2)\epsilon_F} \left(1.13 +  \left(2.6-\frac{\pi^2}{48}\right)\left(\frac{k_BT}{\epsilon_F}\right)^2 + \frac{1}{2}\log\frac{k_BT}{4\epsilon_F}\left(1+\frac{5\pi^2}{12}\left(\frac{k_BT}{\epsilon_F}\right)^2\right) \right)\right),
\end{split}\end{equation}
\end{widetext}
where $P_0 = \frac{3S(S+1)N_b}{8}(1+\alpha)^2\left(\frac{J}{\epsilon_F}\right)^2\left(k_FL\right)\hbar k_F$, and as before, we neglect terms of $O\left(t^0,v^2,\frac{1}{M_b},T^4\right)$.

\bibliography{references}
\end{document}